\documentclass[aps,prl,pdf,superscriptaddress,amsmath,amssymb,amsfonts,twocolumn,showpacs,nofootinbib]{revtex4-2}

\usepackage{amssymb}
\usepackage{eqnarray,amsmath}
\newcommand{\abs}[1]{\left| #1 \right|} 
\usepackage{graphicx}
\usepackage{epsfig}
\usepackage{dcolumn}
\usepackage{bm}
\usepackage{braket}
\usepackage{amsmath}
\usepackage{physics}
\usepackage{csquotes}
\usepackage{mathtools}
\usepackage{graphicx,color,xcolor,colortbl}

\usepackage[colorlinks=true,
linkcolor=blue,
filecolor=blue,      
urlcolor=blue,
citecolor=blue]{hyperref}

\usepackage[normalem]{ulem}

\definecolor{Gray}{gray}{0.85}
\definecolor{LightCyan}{rgb}{0.88,1,1}

\newcolumntype{a}{>{\columncolor{Gray}}c}
\newcolumntype{b}{>{\columncolor{white}}c}

\begin{document}

\title{Quantum Carpets of Higgs particles in a Supersolid}

\author{K. Mukherjee}
\affiliation{Mathematical Physics and NanoLund, Lund University, Box 118, 22100 Lund, Sweden}

\author{M. Schubert}
\affiliation{Mathematical Physics and NanoLund, Lund University, Box 118, 22100 Lund, Sweden}

\author{R. Klemt}
\affiliation{5. Physikalisches Institut and Center for Integrated Quantum Science and Technology, Universität Stuttgart, Pfaffenwaldring 57, 70569 Stuttgart, Germany}

\author{T. Bland}
\affiliation{Mathematical Physics and NanoLund, Lund University, Box 118, 22100 Lund, Sweden}

\author{T. Pfau}
\affiliation{5. Physikalisches Institut and Center for Integrated Quantum Science and Technology, Universität Stuttgart, Pfaffenwaldring 57, 70569 Stuttgart, Germany}

\author{S. M. Reimann}
\affiliation{Mathematical Physics and NanoLund, Lund University, Box 118, 22100 Lund, Sweden}

\begin{abstract}

Supersolids formed from dipolar Bose–Einstein condensates (BECs) exhibit spontaneous density modulation while maintaining global phase coherence. This state of matter supports gapped amplitude (Higgs) excitations featuring a quadratic dispersion relation.  While Higgs modes are typically strongly damped due to coupling with other amplitude and phase modes, imposing an experimentally realistic toroidal geometry allows us to numerically study the time evolution and dispersion of a localized Higgs quasiparticle excitation, with minimal residual coupling to sound modes. Strikingly, the quadratic dispersion leads to the occurrence of (fractional) revivals, similar to those seen in the optical Talbot effect or the so-called quantum carpets. The revival times provide a novel method for determining the effective mass of the Higgs particle through a non-spectroscopic approach. These results pave the way for further studies of coherent Higgs dynamics and mutual interactions between Higgs particles.

\end{abstract}

\date{\today}
\maketitle

According to Goldstone's theorem, breaking a continuous symmetry necessarily gives rise to a massless scalar particle~\cite{Goldstone1962}. In contrast, the massive Higgs amplitude mode---another key excitation at a phase transition---requires additional symmetry protection~\cite{Anderson1963, Higgs1964}. In high-energy physics, this is provided by Lorentz invariance~\cite{Higgs1964}, while in non-relativistic quantum systems, approximate particle-hole symmetry is typically needed~\cite{peker_review_2015}.

Despite this constraint, the Higgs mode has been observed across a wide range of platforms, including superconductors~\cite{Sooryakumar1981, Matsunaga2013, Sherman2015, cea2016, Shimano2020}, quantum antiferromagnets~\cite{Ruegg2008}, and ultracold atomic systems such as fermionic condensates~\cite{Behrle2018, Dyke2024, Kell2024}, bosons near the superfluid–Mott transition~\cite{Bissbort2011, Endres2012}, spinor condensates~\cite{Hoang2016}, and light-coupled Bose gases~\cite{Leonard2017Higgs}. It has also been predicted~\cite{bjerlin2016fewbodyhiggs_predict} and observed~\cite{bayha2020fewbodyhiggs_observe} in few-body fermionic systems. In BCS superfluids, the Higgs mode is tied to the pairing gap and typically exhibits flat dispersion~\cite{Behrle2018}, whereas in bosonic systems with lattices or emergent periodicity, the mode acquires a momentum-dependent dispersion, with the gap closing near a quantum critical point~\cite{Huber2008, Endres2012}.

Experimental detection has so far relied mostly on spectroscopic techniques~\cite{Bissbort2011, Endres2012, Behrle2018, Kell2024, Dyke2024}, such as modulating the pairing gap~\cite{Behrle2018} or the lattice depth~\cite{Endres2012}, and is typically identified through resonant responses. However, spatial inhomogeneities can obscure these signals by coupling amplitude and phase modes~\cite{pollet2012, chen2013, Liu2015}. Directly exciting a single Higgs quasiparticle and tracking its real-space dynamics has remained out of reach, yet such capability would open the door to exploring coherent motion and even Higgs scattering. This challenge is especially compelling in the context of  supersolids, which---once considered elusive~\cite{Boninsegni2012}---have now been realized in several systems~\cite{Li2017, Leonard2017supersolid, Chisholm2024}, including magnetic atoms~\cite{Tanzi2019, Bottcher2019droplet, Chomaz2019, Norcia2021} (see also Refs.\cite{Boettcher2021, Chomaz2023, Recati2023}). These systems support a rich excitation spectrum\cite{Natale2019, Guo2019, Petter2019, Schmidt2021OblateRoton, Hertkorn2021Spectrum2D}, where Higgs excitations coexist with two sound excitations arising from the breaking of distinct continuous symmetries~\cite{Hertkorn2019, Hofmann2021, Ferlaino2023Elongated, sindik2024, Platt2024, hertkorn2024decoupled}.

\begin{figure}
	\centering
	\includegraphics[width = 0.48\textwidth]{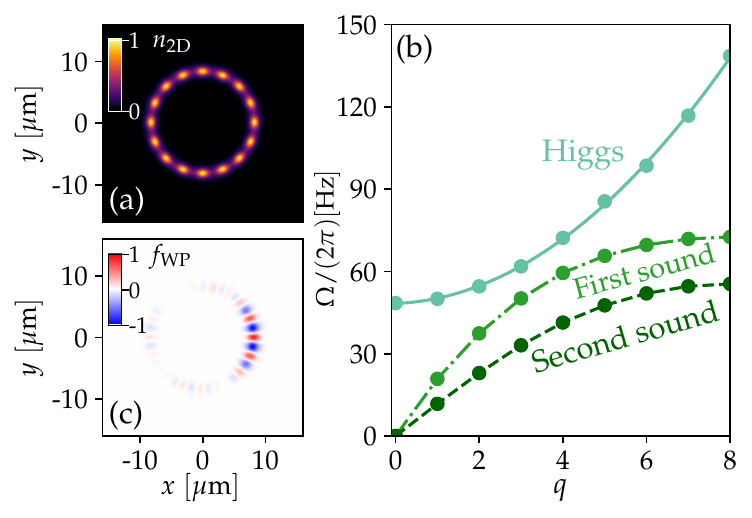}
    \vspace{-0.8cm}
\caption{
Construction of a Higgs wave packet in a toroidal supersolid. 
(a) Normalized column density of an exemplar supersolid state, confined to a toroidal trap with $n_\text{d} = 16$ density peaks. 
(b) First three branches of the elementary excitation spectrum corresponding to the state in (a), plotted as a function of mode number $q$, with the Brillouin-zone edge located at $q = n_\text{d}/2 = 8$. The modes (circles) are connected by lines to guide the eyes.
(c) Real-space structure of a Higgs wave packet, $f_{\text{WP}}$, formed by a linear superposition of the first five Higgs-mode eigenfunctions (see text).
}
\label{figure1}
\end{figure}

In this Letter, we numerically show that a localized Higgs quasiparticle in a dipolar supersolid can be excited in a toroidal trap~\cite{NilssonTengstrand2021, Tengstrand2023, sindik2024}, and that its evolution produces a quantum carpet interference pattern due to the quadratic dispersion and confinement. This enables interferometric detection of the Higgs mode, observation of its time evolution, extraction of its effective mass from revival times, and confirmation of the quadratic dispersion in real space.

Previous work~\cite{hertkorn2024decoupled} demonstrated that in toroidal confinement, amplitude and sound modes decouple across a wide interaction range, even away from criticality. This decoupling allows coherent superposition of multiple Higgs-branch excitations to form a localized wave packet that behaves as a free quantum particle. The resulting dynamics resemble the temporal Talbot effect~\cite{Azana1999, Wen_Talbot_review2013}, where a train of optical pulses re-forms periodically in a dispersive medium, and thus directly probe the Higgs dispersion. Originally observed in optics~\cite{Talbot1836, Rayleigh1881, Berry1996}, Talbot self-imaging has since appeared in plasmonics~\cite{Dennis2007, Li11}, water waves~\cite{Zhang2014, Rozenman2022}, acoustics~\cite{KAIJUN1983295, PRAKASH2000251}, and matter waves~\cite{Clauser1992, chapman1995, Nowak1997, Cahn1997}. In BECs confined to optical lattices, this effect has been used to measure inter-site phase coherence via density self-imaging~\cite{Mark2011, Santra2017, Hoellmer2019, wei2024}. A natural extension is to systems with spontaneously emerging periodicity, such as the dipolar supersolid studied here. Our approach thus offers a novel route to detect Higgs modes and their real-space dispersion, going well beyond standard spectroscopic methods.

\begin{figure}
    \centering
    \includegraphics[width=0.48\textwidth]{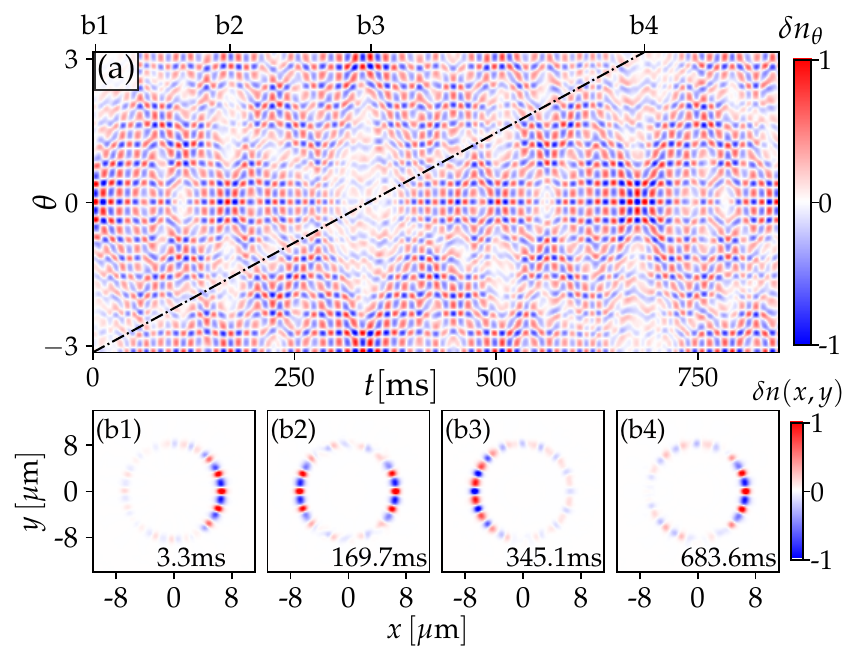}
    \vspace{-0.8cm}
    \caption{Full and partial revivals of Higgs-mode-induced density fluctuations, forming a quantum carpet structure.
    (a) Time evolution of the density fluctuation, $\delta n_\theta$, as a function of the azimuthal angle $\theta$. The dashed line indicates the analytically determined position of a canal, $\theta^{1}_{\rm min}$ (see text). 
    (b1)-(b4) show the normalized density fluctuations at different time instants (highlighted above (a)), corresponding to the full and fractional revivals of the wave packet. Initial state as in Fig.~\ref{figure1}.}
    \label{figure2}
\end{figure}

{\it Theoretical framework}---We consider a dilute, weakly interacting gas of ultra-cold dipolar atoms, where, in the presence of a homogeneous magnetic field, the interatomic interactions are described by the two-body pseudopotential
$U(\mathbf{r}) = \frac{4\pi\hbar^2 a_{\mathrm{s}}}{M} \delta(\mathbf{r}) + \frac{3\hbar^2 a_{\mathrm{dd}}}{M} \frac{1 - 3\cos^2\Theta}{r^3}$, 
where $M$ is the atomic mass, $a_{\mathrm{s}}$ is the tunable $s$-wave scattering length~\cite{Chin2010fri}, and $a_{\mathrm{dd}}$ is the fixed dipolar length. The dipoles are polarized along the $z$-axis, and $\Theta$ is the angle between $\mathbf{r}$ and the polarization direction.

The system is confined in a toroidal trap described by the potential
$V_{\rm trap} = M \omega_r^2 \left[ (r - r_0)^2 + \lambda^2 z^2 \right]/2$,
where $r_0 = 8\,\mu\text{m}$ is the radius of the torus, and $\lambda = \omega_z/\omega_r$ is the aspect ratio. 
For conciseness, in the following we assume parameters corresponding to bosonic $^{162}\mathrm{Dy}$ atoms ($a_{\mathrm{dd}} = 130a_0$, where $a_0$ is the Bohr radius), as well as experimentally realistic trap frequencies of $(\omega_r, \omega_z)/2\pi = (100, 120)\,\text{Hz}$.
 We solve the extended Gross–Pitaevskii equation (eGPE) for the macroscopic wavefunction $\psi(\mathbf{r}, t)$ in imaginary time to obtain the ground state, incorporating the beyond-mean-field Lee-Huang-Yang correction within the local density approximation~\cite{Wachtler2016,Ferrier-Barbut2016,Chomaz2016,Bisset2016}. The interplay between short-range and long-range interactions leads to the emergence of a toroidal supersolid featuring multiple density modulations. To access excitation modes across a wide quasi-momentum range, we illustrate our work in a regime with $n_{\mathrm{d}} = 16$ density peaks at $a_{\mathrm{s}} = 94.5\,a_0$, using a total atom number $N = 1.5 \times 10^5$, as shown in Fig.~\ref{figure1}(a). Note that further details of the model and generalizations of these results to other ring radii are presented in the Supplementary Material~\cite{supmat}.

To compute the excitation spectrum, we linearize the eGPE around the stationary state $\psi_0$ using the ansatz
$
\psi(\mathbf{r}, t) = \left[ \psi_0(\mathbf{r}) + \epsilon \left( u_q(\mathbf{r}) e^{-i \Omega_q t} + v^*_q(\mathbf{r}) e^{i \Omega_q t} \right) \right] e^{-i \mu t / \hbar},
$
which yields the Bogoliubov-de Gennes (BdG) equations. Here, the modes are labled by quantum number $q$, such that $q/r_0$ is the quasi-momentum. Solving these numerically provides the excitation energies $\hbar \Omega_q$ and mode functions $f_q = u_q + v_q$, corresponding to density fluctuations $\delta n_{\text{gs}} = \psi_0 f_q$.

The excitation spectrum of the supersolid ground state is shown in Fig.~\ref{figure1}(b). The superfluid- to supersolid phase transition is driven by a $q=n_\text{d}$ angular roton mode, leading to a periodic density modulation and the Brillouin zone (BZ) structure with  $q \in [-n_\text{d}/2, n_\text{d}/2]$. The first two branches correspond to gapless sound modes that flatten near the BZ edge; the third is a gapped Higgs branch associated with amplitude oscillations.

We extract the effective mass of the Higgs quasiparticle from the curvature of its dispersion near $q = 0$
\begin{align}
\label{MH_BdG}
M_{\rm H} = \frac{\hbar}{4\pi r_0^2} \left(\frac{\partial^2 \Omega}{\partial q^2}\right)^{-1}.
\end{align}
This mass governs the inertial and dispersive properties of a Higgs wave packet: a larger $M_{\rm H}$ implies slower spreading and weaker momentum curvature.

The gap between the sound and Higgs branches enables construction of a localized quasiparticle by superposing BdG modes
$
f_{\text{WP}} = \sum_{q = -\mathcal{M}}^{\mathcal{M}} c_q f^H_q,
$
where $f^H_q$ are Higgs-branch modes and $c_q$ can be found by mapping a Gaussian function onto our basis states (see~\cite{supmat}). An example using five low-$q$ modes ($\mathcal{M} = 5$) is shown in Fig.~\ref{figure1}(c). The resulting fluctuation is peaked near $\theta = 0$, with minor residuals due to the finite system size.

{\it Higgs dispersion and revivals of the wave packet}---To generate a Higgs wave packet in our time-dependent simulations in an experimentally realistic approach, we imprint its spatial profile via a pulsed potential $V_{\text{pulse}}(\mathbf{r}) = \mu f_{\text{WP}}(\mathbf{r})/4$ applied for a brief duration $\tau$. When $\tau$ ($500 ns$ in our work) is much shorter than the timescales of collective modes, the density remains unchanged while the wavefunction acquires a spatially varying phase $\phi(\mathbf{r}) = V_{\text{pulse}}(\mathbf{r})\tau/\hbar$. This phase gradient induces local currents that form a propagating Higgs wave packet. Similar techniques were recently employed in experiments on supersolids~\cite{biagioni2024measurement} and fermionic superfluids~\cite{Del_pace2022}.

The ensuing dynamics are shown in Fig.~\ref{figure2}. In (a), we compute the time-evolved density fluctuation
$
\delta n(\vb{r}, t) = |\psi(\vb{r}, t)|^2 - |\psi(\vb{r}, 0)|^2,
$
and display its azimuthal component, $\delta n_\theta(\theta)$, as a function of angle $\theta$. This immediately reveals interference features and temporally repeating structures in the density. After the application of $V_{\text{pulse}}$, the wave packet becomes fully formed near $t \approx 3.3$ ms, as shown in Fig.~\ref{figure2}(b1), with a near-perfect overlap to the one shown in Fig.~\ref{figure1}(c). In the subsequent dynamics, the wave packet rapidly disperses around the ring, and undergoes self-interference, as well as interference with the underlying supersolid structure. The initial pattern reappears at $t = 683.6$ ms$~\approx T_\text{B}$ [Fig.~\ref{figure2}(b4)]---thus defining full revival time (here referred to as the Talbot time) $T_\text{B}$.

Consistent with this interpretation, we observe localized structures at fractional Talbot times. For example, at $T_\text{B}/2$, the wave packet relocalizes around $\theta = \pi$ [Fig.~\ref{figure2}(b3)], and at $T_\text{B}/4$, it exhibits multiple localized density peaks [Fig.~\ref{figure2}(b3)]. This pattern of full and fractional revivals arises from the underlying quadratic dispersion and persists at long times (not shown here). The resulting $\delta n_{\theta}(\theta, t)$ displays a strikingly ordered structure, with canals of minimum density fluctuation and recurring localization—a quantum carpet reminiscent of the iconic Talbot effect.

\begin{figure}[t]
    \centering
    \includegraphics[width=0.48\textwidth]{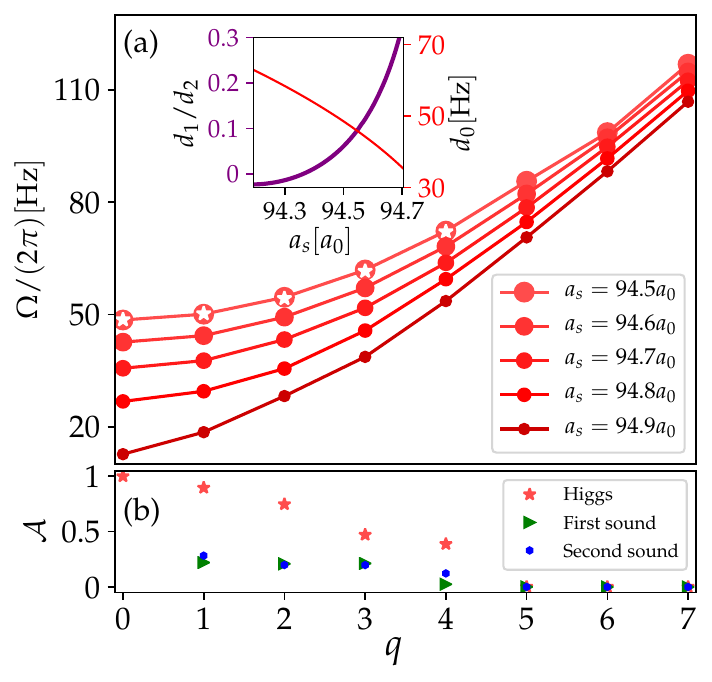}
    \vspace{-0.8cm}
    \caption{Dispersion and spectral weight of Higgs modes.
(a) Higgs-mode dispersion relations from BdG analysis for various scattering lengths $a_s$. Star markers indicate frequencies dynamically extracted at $a_s = 94.5a_0$, showing excellent agreement. The dispersion is well described by a polynomial of second order: $\Omega(q) = d_0 + d_1 \abs{q} + d_2 q^2$.
Inset: Dependence of fit parameters $d_1/d_2$ (left axis) and $d_0$ (right axis) on $a_s$, extracted using the first five Higgs modes.
(b) Normalized spectral weight $\mathcal{A}$ of individual Higgs and sound modes excited by the phase-imprinting protocol at $a_s = 94.5a_0$.}
    \label{figure3}
\end{figure}

Since the wave packet's center does not drift in our system, the group velocity $v_g = \text{d}\Omega/\text{d}q \to 0$, as $q \to 0$. This also ensures symmetric spreading of positive and negative $q$ components. In Fig.~\ref{figure3}(a), we examine the Higgs-mode dispersion across a range of scattering lengths, noting that the density modulation appears for $a_s/a_0 \lesssim 94.92$ under our parameters.

The excitation spectrum becomes increasingly linear near the Brillouin zone edge, with this linearity extending to smaller $q$ as $a_s$ increases. Near the superfluid–supersolid transition, this reflects the softening and eventual closure of the Higgs gap~\cite{Hertkorn2019}. To construct a dispersive wave packet, we focus on momenta away from the BZ edge and interactions deep in the supersolid regime. We quantify the Higgs-mode curvature by fitting the BdG spectrum to $\Omega(q) = d_0 + d_1 |q| + d_2 q^2$, where $d_0$ is the gap at $q = 0$. The inset of Fig.~\ref{figure3}(a) shows how $d_0$ and $d_1/d_2$ evolve with $a_s$, based on the five lowest-$q$ Higgs modes. For $a_s < 94.7,a_0$, the condition $d_1/d_2 < 0.3$ confirms a dominantly quadratic character.

Although our phase-imprinting protocol is designed to selectively excite the Higgs branch, it is important to assess whether sound modes are also inadvertently populated, particularly since Higgs and sound modes at fixed $q$ ($q \geq 1$) belong to the same symmetry class~\cite{hertkorn2024decoupled}. To evaluate this, we compute the overlap between the time-evolved density fluctuation $\delta n(\mathbf{r}, t)$ and the BdG mode functions $f_q$, defined as
 $ O_q(t) = \int \delta n(\vb{r}, t) f_q(\vb{r}) \, \text{d}^3\vb{r} / \sqrt{\int \delta n^2(\vb{r}, t) \, \text{d}^3\vb{r} \int f_q^2(\vb{r}) \, \text{d}^3\vb{r}} $.
We extract the spectral content of each mode via a Fourier transform,
$
A_q(\Omega) = \int O_q(t) e^{i \Omega t} \text{d}t\,,
$
where the maximum amplitude $\mathcal{A} = \max |A_q(\Omega)|$ quantifies the mode’s strength in the dynamics, and the corresponding frequency $\Omega$ gives its characteristic excitation energy.

Figure~\ref{figure3}(b) shows $\mathcal{A}$ for both Higgs and sound modes across various quasi-momenta $q$. The Higgs modes clearly dominate the dynamics, while sound-mode contributions remain below 30\% of their Higgs counterparts at all $q$. This confirms that, under our conditions, the dynamics observed in Fig.~\ref{figure2} are indeed dominated by Higgs-mode contributions.  Since $V_{\rm pulse}$ includes only the first five Higgs mode functions, the $\mathcal{A}$ vanishes for $q > 4$, also highlighting that higgs modes do not couple with each other.

Moreover, the dominant frequencies extracted from $A_q(\Omega)$ at $a_s = 94.5a_0$ closely match the Higgs-mode dispersion obtained from BdG analysis [see star markers in Fig.\ref{figure3}(a)], providing further evidence of the Higgs origin of the quantum carpet patterns in Fig.\ref{figure2}(a). This highlights the precision of our protocol, which excites Higgs modes at their natural frequencies---even though the pulse encodes only spatial information. In contrast to spectroscopic methods based on parametric modulation, which require frequency matching to induce a resonance, our approach relies purely on spatial phase imprinting to selectively initiate amplitude-mode dynamics.

\begin{figure}
	\centering
	\includegraphics[width = 0.48\textwidth]{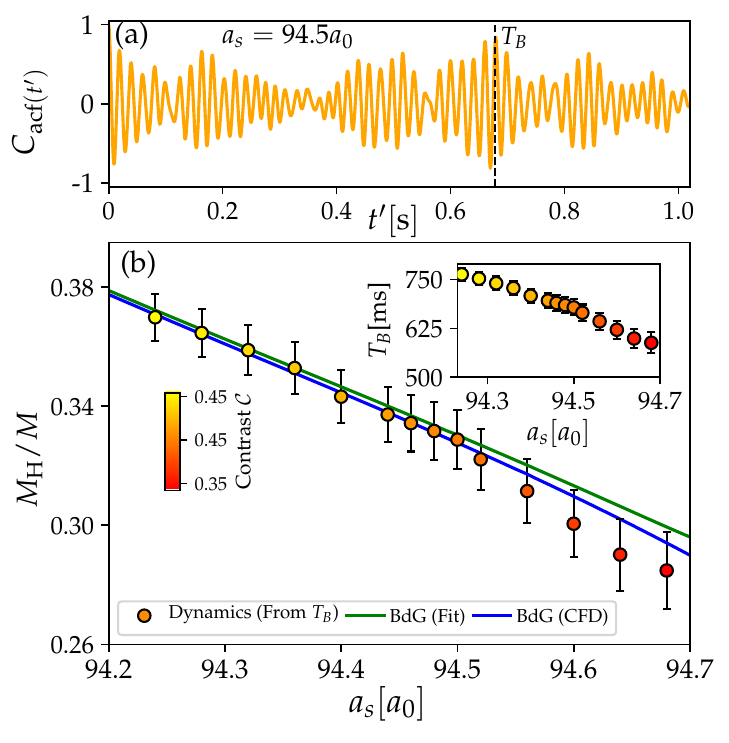}
    \vspace{-0.8cm}
	\caption{Extracting the effective Higgs mass from the Talbot time. (a) Time evolution of the autocorrelation function of the density fluctuations, $C_{\text{acf}}(t')$, at $a_s = 94.5a_0$. The black dashed line marks the Talbot time $T_\text{B}$. (b) Effective Higgs mass $M_\text{H}$ as a function of $a_s$, extracted via three methods: quadratic fit to the BdG dispersion (green line), central finite difference (CFD, blue line), and from real-time dynamics (colored markers) by measuring $T_\text{B}$ (see inset), derived from $C_{\text{acf}}$. Error bars reflect the uncertainty associated with the fast oscillation scale $1/d_0$. Marker shading encodes the density contrast $\mathcal{C}$ of the corresponding ground state.}
	\label{figure4}
\end{figure}

{\it Higgs mass from the revival times}---To gain insight into the quantum carpet structure, we approximate the profiles of the Higgs-mode functions along the azimuthal directions as $f^{H}_{q} (\theta) \propto \cos(n_d \theta) \cos(q\theta)$. The total wave packet takes the form $f_{\rm WP}(\theta) \approx \mathcal{N} \cos(n_d \theta) \Lambda(\theta, 0),$ with $\Lambda(\theta, t)$, describing the envelope of the azimuthal density fluctuation~\cite{supmat}, and  $\mathcal{N}$ is a normalization factor. The evolution of $\Lambda(\theta, t)$ is governed by an effective Schr\"odinger equation. From this, the time-dependent fluctuation follows~\cite{supmat}
\begin{equation}
\label{densfluc}
\delta n(\theta,t) \propto \psi_0\cos (n_d\theta)\sum_qc_q\cos\left[ q\theta+\frac{\hbar q^2}{2M_Hr_0^2}t\right]
\end{equation}
revealing that the pattern arises from quadratic dispersion relation and closed confinement.
The positions of canals visible in Fig.~\ref{figure2}(a), are determined by
\begin{equation}
\label{theta_min}
\theta^{q}_{\min} = \pm \left( \frac{t \hbar q}{2 M_\text{H} r_0^2} + \frac{\pi}{q} \right).
\end{equation}
When the canals contributed by the two modes coincide with each other, partial revivals takes place, with the corresponding  revival time being determined by
\begin{equation}
    t^{q{_1},q{_2}}_{r} = -\frac{2 \pi  M_\text{H}r^2_{0}}{q_1 q_2 \hbar}.
\end{equation}
 Here, $q_1$ and $q_2$ have opposite sign, featuring two counter propagating modes. Note that the wave packet fully revives at every $T_\text{B}/2$, where $T_\text{B} = 4 \pi M_{\rm H} r_0^2/\hbar$, with the revival at its original position occurring at $T_\text{B}$. The above discussion is also consistent with the observation in Fig.~\ref{figure2}, see also Ref.~\cite{supmat}. This analytic result links the revival dynamics to the effective mass of the Higgs quasiparticle. 
We thus see that, importantly, either measuring full  or partial  revival times---i.e., dynamically tracking the positions of canals in the density fluctuations---provides a direct probe of $M_\text{H}$, representing the central result of this work. 

With these insights, we propose extracting the Higgs mass experimentally by measuring the Talbot time via the autocorrelation function of the density fluctuations:
\begin{equation}
C_{\mathrm{acf}}(t') = \int \text{d}t\, \frac{\int \delta n(\mathbf{r}, t)\, \delta n(\mathbf{r}, t + t')\, \text{d}^3\mathbf{r}}{\int \delta n^2(\mathbf{r}, t)\, \text{d}^3\mathbf{r}}\,.
\end{equation}
This function quantifies the temporal self-similarity of the density fluctuations, with peaks in $|C_{\mathrm{acf}}(t')|$ signaling Talbot revivals. The full Talbot time $T_\text{B}$ can be extracted from the position of the dominant revival peak, with an uncertainty of order $1/d_0$, where $d_0$ is the Higgs-mode gap. In Fig.~\ref{figure4}(a), we show $C_{\mathrm{acf}}(t')$ for $a_s = 94.5\,a_0$, with the Talbot revival marked by a dashed line. Scanning across $a_s$, we then compute the effective mass of the Higgs quasiparticle from the measured Talbot time, as shown in Fig.~\ref{figure4}(b).
We observe an increase in mass with decreasing scattering length, consistent with predictions from $O(N)$ field theories, where the Higgs mass similarly evolves with the critical parameter~\cite{Podolsky2011, Gazit2013}.
Furthermore, a correspondence between this behavior and the ground state structure can be established by calculating the density contrast $\mathcal{C} = (n_{\rm max} - n_{\rm min}) / (n_{\rm max} + n_{\rm min})$. Here $n_{\rm max}$ and $n_{\rm min}$ are the local maximum and minimum densities, respectively. This contrast quantifies the strength of spontaneous density modulation, interpolating from $\mathcal{C} \to 0$ in the unmodulated (superfluid) limit to $\mathcal{C} \to 1$ in a fully modulated (crystalline) limit, analogous to the behavior of the superfluid fraction~\cite{Tanzi2019, Bottcher2019, Chomaz2019}. As $\mathcal{C}$ increases, the Higgs quasiparticle becomes heavier, resulting in longer revival times.

Finally, we compare the Higgs mass extracted dynamically with that obtained from the excitation spectrum via Eq.~\eqref{MH_BdG}, using both finite differences between discrete BdG modes and a least-squares fit to a quadratic dispersion. As shown in Fig.~\ref{figure4}(b), the two approaches agree remarkably well, deviating only when the dispersion ceases to be quadratic [Fig.~\ref{figure3}(b)].  We also validate the robustness of our approach by examining Talbot dynamics and Higgs mass extraction across different numbers of density sites, as detailed in the Supplemental Material~\cite{supmat}.

{\it Summary}---We have introduced an interferometric scheme based on the temporal Talbot effect to investigate the Higgs modes in a dipolar supersolid. Unlike the spectroscopic techniques, this method reveals signatures of amplitude modes directly in real space. Using toroidal confinement to decouple amplitude and phase modes, we construct a localized Higgs wave packet from low-lying excitations exhibiting a well-defined quadratic dispersion. This dispersion directly manifests in the partial revivals of the wave packet after a phase-imprinting pulse initiates the dynamics.
The effective mass of the Higgs particle is directly encoded in the Talbot time, which can be extracted from the autocorrelation of the evolving density fluctuations.

This approach offers a powerful alternative to spectroscopic methods for detecting amplitude modes in quantum many-body systems. By linking the  dynamics of a collective excitation to its effective mass, our work opens a new route to precision measurements of Higgs-like modes. Looking forward to future experiments, this protocol may inspire interferometric detection schemes in dipolar supersolids, and potentially in other platforms such as light-mediated supersolids~\cite{Li2017, Leonard2017, Chisholm2024}.

{\it Acknowledgments}---We are grateful to J.~Hertkorn, P.~St\"urmer, S.~Welte, F.~Hellstern, P.~Uerlings, T.A.~Cardinale, L.~Chergui, D.~Gaur  for insightful conversations. K.M., M.S., T.B. and S.M.R. thank the Knut and Alice Wallenberg Foundation (Grant No. KAW 2018.0217) and the Swedish Research Council (Grant No. 2022-03654VR) for financial support. R.K and T.P. acknowledge funding from the European Research Council (ERC) (Grant Agreement No. 101019739).

\bibliography{reference}

% Optional: page break

\newpage

\section{Supplementary Material}

\subsection{Construction of a Higgs wave packet}

Here, we detail the construction of the wave packet $f_{\mathrm{WP}}$. Our objective is to design a function that approximates a localized wave packet confined to a small number of density sites. As a starting point, we consider the $q = 0$ Higgs mode, $f^\text{H}_{0}$, which exhibits alternating signs on the density sites and in the interstitial regions, see Fig.~\ref{sup_fig1}(a). To localize this mode on just a few sites, we multiply it by a Gaussian envelope, yielding the target function
\begin{equation}
    f_{\mathrm{target}} = f^\text{H}_{0}\, e^{-\theta^2/w^2},
\end{equation}
where we choose $w = 0.5$ in this work. This value of $w$ ensures a reasonably localized wave packet.

To construct the corresponding superposition of Higgs modes, we project $f_{\mathrm{target}}$ onto the full set of orthonormal mode functions $\{f_q^\text{H}\}$ by evaluating the projection coefficients
\begin{equation}
    c_q = \frac{\int\text{d}^3\vb{r}\, f_{\mathrm{target}}\, f^\text{H}_{q}}{\int \text{d}^3\vb{r}\,\left(f^\text{H}_{q}\right)^2}.
\end{equation}
The localized wave packet is then reconstructed via
\begin{equation}
    f_{\mathrm{WP}} = \sum_q c_q f^\text{H}_{q}.
\end{equation}
Note that only in the limit of an infinite number of contributing modes would $f_{\mathrm{WP}}$ exactly reproduce $f_{\mathrm{target}}$. Yet, even in our finite system, the resulting $f_{\mathrm{WP}}$ remains well-localized and serves as an excellent approximation to the desired target wave function, as we have shown in the main text.

We remark that a similar procedure can be applied to construct localized wave packets in momentum space. An example of such a momentum-space packet is shown in Fig.~\ref{sup_fig1}(b), at $a_s = 95.5\,a_0$, with all other parameters identical to those used in the main text.

\subsection{Analytical insights}
In this section, we showcase that the quantum carpet patterns realized by the experimentally feasible pulse potential in the main text can also be understood from an insightful analytical model. After integrating over $z$ and $r$, we write the initial wave packet as 
\begin{align}
\label{Eq1_SM}
    f_{\rm WP} = \mathcal{N} \cos(n_\text{d}\theta) \sum_{q=-\mathcal{M}}^\mathcal{M} c_q \cos(q\theta) \approx \mathcal{N} \cos(n_\text{d}\theta)\Lambda(\theta,t=0)\, .
\end{align}
In the limit $\mathcal{M} \rightarrow \infty$, the sum becomes a full Fourier series, allowing for the construction of well-localized, symmetric wave packets of any desired shape $\Lambda(\theta, t = 0)$. By adjusting the constant $\mathcal{N}$, one can match the peak density fluctuation of the analytical wave packet with that obtained from the pulse potential. However, the full and partial revivals of the wave packet do not depend on $\mathcal{N}$.

\begin{figure}
    \centering
    \includegraphics[width=0.48\textwidth]{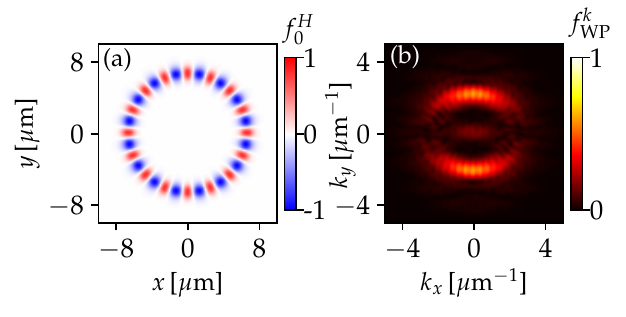}
    \vspace{-0.8cm}
    \caption{
        (a) Mode function $f^\text{H}_{0}$ corresponding to the $q = 0$ Higgs mode, and (b) a momentum-space density $f^{k}_{\text{WP}}$ representing a wave packet constructed in momentum space. The scattering length is fixed at $a_s = 94.5\,a_{0}$, and all other parameters are the same as in the main text.
    }
    \label{sup_fig1}
\end{figure}

In this work, we focus on constructing a Gaussian wave packet given by $\Lambda = \exp(-\theta^2 / w^2)$. Notably, the term $\cos(n_\text{d} \theta)$ emulates the profile of the $q = 0$ Higgs mode, as shown in Fig.~\ref{sup_fig1}(a). As demonstrated earlier, the coefficients $c_q$ can be found by mapping $\Lambda$ onto our basis states:
\begin{align}
    c_q = \frac{1}{2\pi} \int \text{d}\theta\, \Lambda(\theta) \cos(q\theta)\, = \frac{w}{2 \sqrt{\pi}}\exp\left(-\frac{q^2 w^2}{4} \right)\, ,
\end{align}
which is valid for $w\ll \pi$. Due to the finite number of localized density sites (sixteen in our case), the number of basis states is $\mathcal{M} \leq 8$. Consequently, our wave packet $f_{\rm WP}$ deviates from a perfect Gaussian shape.

As explicated in the main text, the Higgs-mode dispersion relation takes the form $\Omega(q) = d_2 q^2 + d_1 |q| + d_0$. For scattering lengths satisfying $d_2 \gg d_1$, the dynamics of the wave packet can be effectively described by the free-particle Schrödinger equation in momentum space
\begin{equation}
    i \hbar \frac{d \Lambda}{dt} = d_2 k_\theta^2 \Lambda = \frac{\hbar^2 k_\theta^2}{2 M_\text{H} r_0^2} \Lambda\,,
\end{equation}
where $\hbar k_\theta$ is the angular momentum. An energy rescaling allows us to disregard the constant term $d_0$. The coefficient $d_2$, using Eq.~(1) from the main text, can be directly related to the Higgs mass $M_\text{H}$. We can find the time evolution by applying the time-evolution operator onto $\Lambda(k_\theta,t=0) = \sum_q c_q\left[ \delta(k_\theta - q)+\delta(k_\theta + q)\right]/2$
\begin{align}
    \Lambda(k_\theta, t) = \exp\left(-i\frac{\hbar k_\theta^2}{2 M_\text{H} r_0^2} t \right) \Lambda(k_\theta,t=0)\, .
\end{align}
Transforming back into real space gives
\begin{align}
\Lambda(\theta, t) =  \sum_{q=-\mathcal{M}}^\mathcal{M} c_q \cos\left(  q \theta +  \frac{\hbar q^2}{2 M_\text{H} r_0^2} t \right)\, .        
\end{align}
The wavefunction can then be constructed by
\begin{align}
    \psi_\theta(t) \approx e^{-i\mu t/\hbar} \left[\sqrt{n_{\theta}(\theta, t = 0)} + \mathcal{N} \cos(n_\text{d}\theta) \Lambda(\theta, t)\right]\, .
\end{align}
Then, defining the density fluctuation as $\delta n_{\theta}(\theta, t) = n_{\theta}(t) - n_{\theta}(t = 0)$, we arrive at
\begin{align}
\delta n_{\theta}(\theta, t)& =\ 
 2\mathcal{N} \sqrt{n_{\theta}} \cos(n_\text{d} \theta) 
\sum_q c_q \cos\left( q\theta + \frac{\hbar q^2}{2 M_\text{H} r_0^2} t \right) \nonumber \\
&\notag + \mathcal{N}^2 \cos^2(n_\text{d} \theta)  \sum_{q_1,q_2} c_{q_1}c_{q_2} \cos\left( q_1\theta + \frac{\hbar q_1^2}{2 M_\text{H} r_0^2} t \right)\\ 
&\times\cos\left( q_2\theta + \frac{\hbar q_2^2}{2 M_\text{H} r_0^2} t \right) \,.
\end{align}

\begin{figure}
    \centering
    \includegraphics[width=0.48\textwidth]{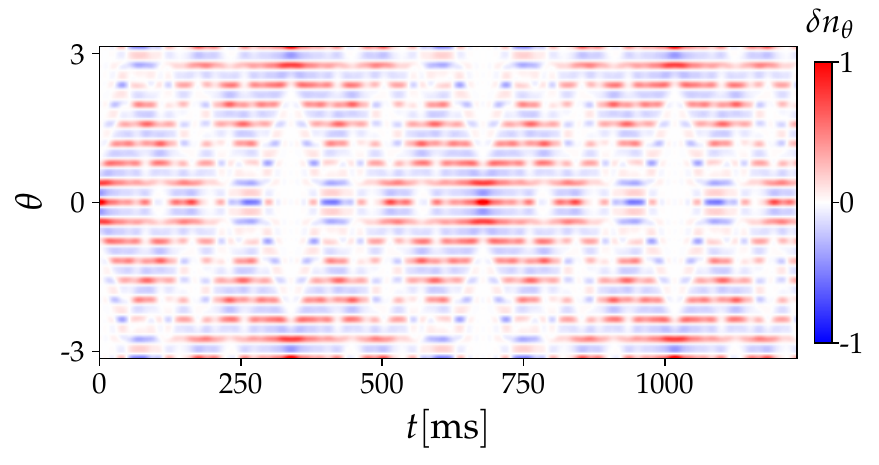}
    \vspace{-0.8cm}
    \caption{Talbot interference dynamics and fractional revivals of the density fluctuations calculated from the analytical formula (see main text), for a scattering length of $a_s = 94.5a_0$.}
    \label{figure2_sup}
\end{figure}

Note that the above expression captures both interference among Higgs modes and interference between each Higgs mode and the periodic supersolid background. However, we emphasize that the emergence of the quantum carpet structure fundamentally arises from the quadratic dispersion relation and the discreteness of the mode number $q$, which reflects the finite, confined geometry.
While interference is inevitable, it only affects the magnitude of $\delta n_{\theta}(\theta, t)$ and does not alter the fundamental structure of the pattern.
We neglect terms of order $\mathcal{O}(\mathcal{N}^2)$, as the normalization factor $\mathcal{N}$ is typically small, and retain only the leading-order contribution. This leads directly to Eq.~(2) of the main text.

Each mode can contribute to the minimum of the density fluctuation $\delta n_{\theta}$ (the so-called canals). They can be found by setting
\begin{equation}
    q \theta^{q}_{\rm min} + \frac{\hbar q^2}{2 M_\text{H} r_0^2} t = \pm \pi\,.
\end{equation}
If the canals contributed by momenta $q_{1}$ and $q_{2}$ coincide with each other, then we have
\begin{equation}
    \theta^{q_1}_{\rm min} = \theta^{q_2}_{\rm min}\,.
\end{equation}
The above condition determines the partial revival times and leads to Eq.~(4) in the main text.

The resulting density fluctuation $\delta n_{\theta}(\theta, t)$, derived from the analytical expression, is shown in Fig.~\ref{figure2_sup} for a scattering length of $a_s = 94.5a_0$. For a Gaussian wave packet, the absolute value of each coefficient $c_q$ decreases with increasing mode number $q$, such that only the lowest few modes significantly contribute to the dynamics. The $q = 0$ Higgs mode does not contribute to the formation of canals. The dominant contribution arises from the $q = \pm1$ mode, which produces a pronounced line at $\theta^1_{\rm min}$ (see also the dashed line in Fig.~2 of the main text for $q = -1$). Additional canals associated with $q = \pm2$ modes are also visible in Fig.~\ref{figure2_sup}. In each time interval $\left(t, t + T_\text{B}\right)$, we expect $2q_1q_2$ fractional revivals between modes with mode numbers $q_1$ and $q_2$. However, for large $q_i$, the fractional revivals are faintly visible.

Both full and partial revivals of the wave packet---signatures of the underlying quadratic dispersion---are thus clearly observed and are in excellent agreement with the dynamics shown in Fig.~2. This close correspondence further confirms the efficacy of the pulsed potential technique as a robust and experimentally accessible method for exciting massive Higgs quasiparticles.

\subsection{Quantum carpets for varying number of localized density sites}

\begin{figure}[t]
    \centering
    \includegraphics[width=0.48\textwidth]{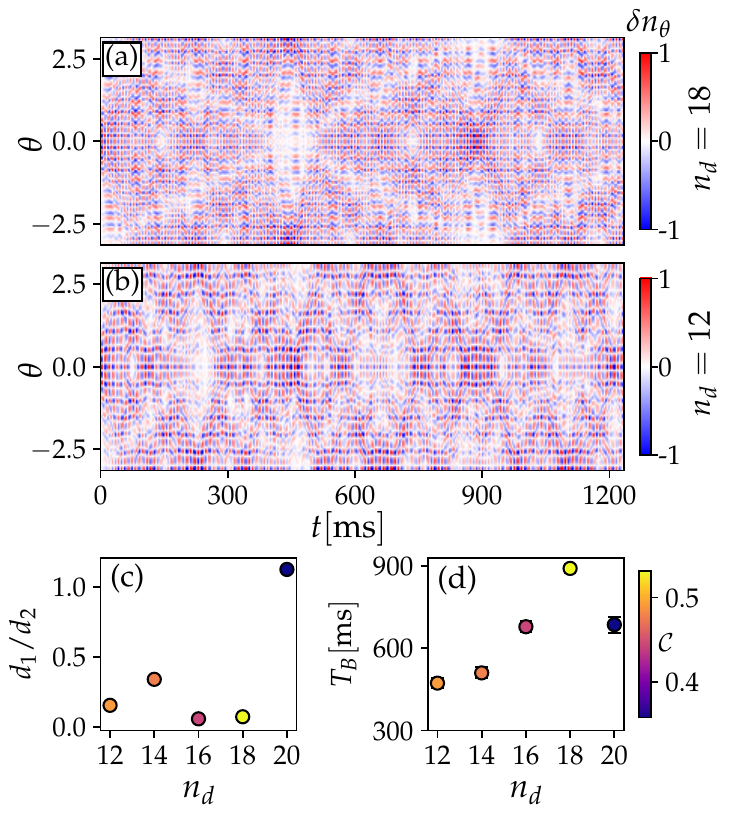}
    \vspace{-0.8cm}
    \caption{
Talbot interference for varying numbers of density sites. 
(a, b) The density fluctuation $ \delta n_{\theta}(\theta) $ for $n_\text{d} = 18$ and $n_\text{d} = 12$ sites, respectively. 
(c) Ratio of dispersion coefficients $d_1/d_2$ and (d) Talbot time $T_\text{B}$ versus $n_\text{d}$. 
Color indicates the ground-state contrast $\mathcal{C}$, with matching color-coding in (c) and (d). 
The scattering length is fixed at $a_s = 94.5\,a_0$, while $N$ and $r_0$ are adjusted to vary $n_\text{d}$ (see text). 
Dynamics are initiated by a pulse potential as described in the main text.
    }
    \label{sup_varying_sites}
\end{figure}

In the main text, we analyzed Talbot interferometric dynamics for a system with $n_\text{d} = 16$ localized density sites. Here, we demonstrate the generality of these findings by extending the same analysis to systems with varying numbers of density sites. The number of density sites in the system can be tuned by adjusting several free parameters. In particular, we fix the scattering length at $a_s = 94.5\,a_0$, retain the trapping frequencies used in the main text, and systematically vary the total atom number $N$ and the radial system size $r_0$, while keeping the ratio $N/r_0$ constant. This fixes the linear density of the system, which, in the ideal scenario of large ring radii---where the system can be approximated as an infinite linear chain---determines the quality of the supersolid~\cite{Blakie2020supersolelongate,smith2023supersolidity}. This controlled scaling yields configurations with $n_\text{d} = 20$, $18$, $14$, and $12$ density sites for $(N, r_0) = (187500,\ 10 \mu \text{m})$, $(168750,\ 9  \mu \text{m})$, $(131250,\ 7 \mu \text{m})$, and $(112500,\ 6 \mu \text{m})$, respectively.

For each configuration with a different number of density sites, the critical point of the SF–SS phase transition shifts accordingly. As a result, the relative proximity of the fixed scattering length $a_s = 94.5\,a_0$ to the corresponding critical point depends on $n_\text{d}$, which in turn affects the density contrast in the ground state. To illustrate this behavior, we present the time evolution of the density fluctuation $\delta n_{\theta}$ for $n_\text{d} = 18$ and $n_\text{d} = 12$, shown in Figs.~\ref{sup_varying_sites}(a) and Figs.~\ref{sup_varying_sites}(b) respectively. The interference dynamics and revival of the wave packet are more pronounced for $n_\text{d} = 18$ than for $n_\text{d} = 12$, indicating a more dominant quadratic component in the dispersion relation and a ground-state configuration located further from the critical point.

This interpretation is further supported by analyzing the ratio of the linear to quadratic dispersion coefficients, $d_1/d_2$, along with the ground-state density contrast $\mathcal{C}$; see Fig.~\ref{sup_varying_sites}(c). For $n_\text{d} = 18$, we find $d_1/d_2 \approx 0.07$ and $\mathcal{C} \approx 0.53$, whereas for $n_\text{d} = 12$, the values are $d_1/d_2 \approx 0.16$ and $\mathcal{C} \approx 0.49$. The Talbot time $T_\text{B}$ for systems with different numbers of density sites is shown in Fig.~\ref{sup_varying_sites}(d). Both $T_\text{B}$ and the effective mass of the Higgs wave packet (not shown) reach their maximum values when $d_1/d_2$ is small and the ground state exhibits high density contrast---conditions that are clearly reflected in the observed interference dynamics. However, as the system approaches the phase transition, this signature becomes less distinct due to the increasing contribution of the linear dispersion component. For example, in the case of $n_\text{d} = 20$, where $\mathcal{C} \approx 0.35$, the interference patterns are noticeably less pronounced (not shown), as in $n_\text{d} = 12$ case.

In summary, by varying $n_\text{d}$ we can deduce that a quadratic dispersion relation emerges for the Higgs excitation branch when the system lies sufficiently far from the critical point. This regime is marked by a high density contrast in the corresponding SS ground state. Under these conditions, an initially localized wave packet exhibits prominent Talbot interference dynamics, which can be exploited to accurately extract the effective mass of the Higgs mode.

\subsection{Details of the numerical simulations}

Here we detail the numerical methods used to obtain the results in both the main text and Supplemental Material. Simulations are performed at zero temperature within the mean-field framework of the extended Gross-Pitaevskii equation (eGPE), given by~\cite{Wachtler2016,Ferrier-Barbut2016,Chomaz2016,Bisset2016}
\begin{equation}
i\hbar \frac{\partial \psi}{\partial t} = \mathcal{L} \psi\,,
\end{equation}
where the eGPE operator takes the form
\begin{align}
\label{eGPE}
\mathcal{L} = & -\frac{\hbar^2}{2M}\nabla^2 + V(\vb{r}) + g |\psi(\vb{r},t)|^2 +  \gamma(\epsilon_{\mathrm{dd}}) |\psi(\vb{r},t)|^3 \nonumber\\ &+ \frac{3}{4\pi} g_{\mathrm{dd}} \int \text{d}^3\vb{r}'\, \frac{1 - 3\cos^2\Theta}{|\vb{r} - \vb{r}'|^3} |\psi(\vb{r}',t)|^2 \,.
\end{align}
The potential $V(\vb{r})$ includes both the static toroidal confinement $V_{\rm trap}(\vb{r})$ and the time-dependent pulse potential $V_{\rm pulse}(\vb{r})$, as defined in the main text, used to initiate wave-packet dynamics. The contact interaction strength is given by $g = 4\pi \hbar^2 a / M$, where $a$ is the tunable $s$-wave scattering length. The dipolar interaction term depends on the angle $\Theta$ between $\vb{r} - \vb{r}'$ and the $z$-axis, along which the dipoles are polarized. The dipolar coupling is $g_{\mathrm{dd}} = 4\pi\hbar^2 a_{\mathrm{dd}} / M$, with $a_{\mathrm{dd}} = \mu_0 \mu_m^2 M / (12\pi \hbar^2)$ the dipolar length. The last term in Eq.~\eqref{eGPE} is the Lee–Huang–Yang (LHY) correction, with~\cite{Lee1957,Schutzhold2006,Lima2011}
\begin{equation}
\gamma = \frac{32}{3}g \sqrt{\frac{a^3}{\pi}}\left(1 + \frac{3}{2} \epsilon_{\mathrm{dd}}^2\right), \quad \epsilon_{\mathrm{dd}} = a_{\mathrm{dd}} / a\,.
\end{equation}
For numerical implementation, the eGPE is rescaled to dimensionless form using characteristic length and time scales $l_s$ and $t_s = M l_s^2 / \hbar$. The ground state is obtained via imaginary-time propagation using a split-step Fourier spectral method. Real-time evolution is then used to study dynamical behavior. The nonlocal dipolar potential is evaluated efficiently in momentum space with a spherical cutoff equal to half the simulation box size to avoid aliasing effects from periodic boundary conditions.

To analyze collective excitations, we solve the Bogoliubov–de Gennes (BdG) equations. This is done by linearizing the eGPE around the stationary solution $\psi_0$ via the ansatz
\begin{equation}
\psi(\vb{r},t) = \left[ \psi_0(\vb{r}) + \epsilon \left( u(\vb{r}) e^{-i\Omega t} + v^*(\vb{r}) e^{i\Omega t} \right) \right] e^{-i\mu t/\hbar}.
\end{equation}
Substituting into Eq.~\eqref{eGPE} and retaining terms linear in $\epsilon$ yields the eigenvalue problem
\begin{equation}
\label{bdg}
\begin{pmatrix}
\mathcal{L} - \mu + X & X \\
- X & -(\mathcal{L} - \mu + X)
\end{pmatrix}
\begin{pmatrix}
u \\
v
\end{pmatrix}
= \hbar \Omega 
\begin{pmatrix}
u \\
v
\end{pmatrix},
\end{equation}
where the operator $X$ acts on a function $q = u, v$ as
\begin{align}
X q(\vb{r}) = & \int \text{d}^3\vb{r}'\, U_{\mathrm{dd}}(\vb{r} - \vb{r}')\, \psi_0(\vb{r}') \psi_0(\vb{r}) q(\vb{r}') \nonumber \\
& + g |\psi_0(\vb{r})|^2 q(\vb{r}) + \frac{3}{2}\gamma |\psi_0(\vb{r})|^3 q(\vb{r}).
\end{align}
To simplify numerical diagonalization, we switch to symmetric and antisymmetric combinations of the Bogoliubov amplitudes, $f = (u + v)/\sqrt{2}$ and $g = (u - v)/\sqrt{2}$, reducing the BdG problem to coupled equations with halved dimensionality. These are solved using standard sparse eigenvalue solvers.

Simulations are carried out in a three-dimensional box with grid resolutions of \(128 \times 128 \times 64\) and \(256 \times 256 \times 64\), confirming convergence of BdG excitation frequencies. Sensitivity to grid resolution is minimal; instead, accurate results depend primarily on the physical box size, which we fix to \(L_x = L_y = L_z = 45\,\mu\mathrm{m}\) to prevent interaction with periodic images. The spatial grid spacing is $\Delta_i = L_i / n_i$ for $i = x, y, z$, and the time step used in time evolution is $\Delta t = 10^{-4}$.

\end{document}